\documentclass[
reprint,
superscriptaddress,
amsmath,amssymb,
aps,
prl,
]{revtex4-2}

\usepackage{graphicx}  
\usepackage{dcolumn}  
\usepackage{bm}       
\usepackage[colorlinks=true, allcolors=blue]{hyperref}  
\usepackage{physics}

\begin{document}
	\title{Cavity-Mediated Gas-Liquid Transition}
	
	\author{Fan Zhang}
	\affiliation{Hefei National Laboratory, University of Science and Technology of China, Hefei 230088, China}
	\author{Haowei Li}
	\affiliation{Laboratory of Quantum Information, University of Science and Technology of China, Hefei 230026, China}
	\author{Wei Yi}
	\email{wyiz@ustc.edu.cn}
	\affiliation{Laboratory of Quantum Information, University of Science and Technology of China, Hefei 230026, China}
	\affiliation{Anhui Province Key Laboratory of Quantum Network, University of Science and Technology of China, Hefei, 230026, China}
	\affiliation{CAS Center For Excellence in Quantum Information and Quantum Physics, Hefei 230026, China}
	\affiliation{Hefei National Laboratory, University of Science and Technology of China, Hefei 230088, China}
	\affiliation{Anhui Center for Fundamental Sciences in Theoretical Physics, University of Science and Technology of China, Hefei 230026 China}

	\begin{abstract}
		We study the gas-liquid transition in a binary Bose-Einstein condensate, where the two Zeeman-shifted hyperfine spin components are coupled by cavity-assisted Raman processes.
		Below a critical Zeeman field, the cavity becomes superradiant for an infinitesimally small pumping strength, where the enhanced superradiance is facilitated by the simultaneous formation of quantum droplet, a self-bound liquid phase stabilized by quantum fluctuations.
		Above the critical Zeeman field, the gas-liquid transition only takes place at a finite pumping strength after the system becomes superradiant.
		As the back action of the gas-liquid transition, the superradiant cavity field undergoes an abrupt jump at the first-order transition point.
		Furthermore, as a result of the fixed density ratio of the quantum droplet, the cavity field exhibits a linear scaling with the pumping strength in the liquid phase.
		These features serve as prominent signals for the cavity-mediated gas-liquid transition and coexistence, which derive from the interplay of Zeeman field, cavity-assisted spin mixing, and quantum fluctuations.
	\end{abstract}

	\maketitle
	{\it Introduction.---}
	The observation of quantum droplets in dilute gases of dipolar or binary Bose-Einstein condensates have enriched our understanding of quantum matter~\cite{petrov2015quantum,chomaz2016quantum,schmitt2016self,ferrier2016observation,bottcher2019transient,semeghini2018self,derrico2019observation,guo2021lee,chomaz2019long,cabrera2018quantum,cheiney2018bright,boudjemaa2018fluctuations,bisset2021quantum,smith2021quantum,gallemi2022superfluid,tanzi2019observation}.
	These droplets are an exotic self-bound quantum fluid, stabilized by beyond-mean-field quantum fluctuations~\cite{petrov2015quantum,gu2020phonon,zhang2022phonon,2021xiong,zhang2025density}. In recent years, their origin~\cite{petrov2015quantum,hu2020consistent,wang2020theory}, as well as the accompanying gas-liquid transition and quantum criticality~\cite{he2023quantum,gu2023liquid,mithun2020modulational,flynn2023quantum,tengstrand2022droplet}, have stimulated persistent interest, which further inspires ingenious schemes to engineer exotic states.
	For instance, in a series of recent experiments, droplet crystals in dipolar condensates have emerged as a versatile platform for the study of supersolids~\cite{tanzi2019supersolid,guo2019low,tanzi2019observation,buhler2023quantum,wenzel2017striped,hertkorn2021pattern,ripley2023two,gallemi2022superfluid,zhang2019supersolidity}, offering valuable insights to their unique properties.
	On the other hand, it is shown that, by coupling the two spin components or adjusting the population imbalance in a box potential, one is able to tune the gas-liquid transition and coexistence in a binary Bose-Einstein condensate~\cite{gu2023liquid,he2023quantum}.
	While observing and tuning gas-liquid coexistence is crucial for the study of the quantum criticality therein, an outstanding problem is the identification of droplet formation that heralds the gas-liquid coexistence regime. The discontinuity of the density profile typical of the liquid-gas coexistence under the local density approximation is easily smoothed out by the trapping potential, making it difficult to serve as a sensitive signal.

	In this work, we address the issue by proposing a configuration where the two Zeeman-shifted hyperfine spin components of a binary Bose-Einstein condensate are coupled by cavity-assisted Raman processes.
	As is the case with typical atom-cavity hybrid systems~\cite{hepp1973superradiant,wang1973phase,larson2017some,nagy2010dicke,chen2014superradiance,keeling2014fermionic,piazza2014umklapp,mivehvar2017disorder,baumann2010dicke,zhang2021observation}, while the cavity field actively drives the gas-liquid transition by concocting (with the Zeeman field) the desirable density ratio of spins, the back action of the transition gives rise to unique signatures in the cavity field.
	Thus, our hybrid system not only provides useful signals for detection, but also serves as an intriguing platform for studying the gas-liquid transition under a dynamic cavity coupling.
	Starting from a fully spin-polarized gas, we show that, the cavity-assisted Raman coupling
	competes against the Zeeman field, enabling spin mixing that is crucial for both the superradiance and gas-liquid transition.
	Specifically, below a critical Zeeman field, the superradiance is dramatically enhanced and occurs simultaneously with the quantum-droplet formation at an infinitesimally small pumping strength.
	Whereas above the critical Zeeman field, increasing the pumping strength only sequentially triggers superradiance and gas-liquid transition.
	Based on such a picture, we analytically derive the critical Zeeman field, which is then confirmed through self-consistent numerical calculations.
	Importantly, at the gas-liquid transition, the sharp change in the density profile accompanying the droplet formation leads to an abrupt jump in the superradiant cavity field.
	Further, in the liquid phase, the fixed density ratio of the quantum droplet gives rise to a linear scaling of the cavity field with the pumping strength.
	While these detectable features unambiguously mark the gas-liquid transition and coexistence in our configuration, we further map out the phase diagram of the system, thus paving the way for future studies of the rich quantum criticality herein.

	\begin{figure}[tbp]
		\centering

		\begin{minipage}[t]{0.48\linewidth}
			\vspace*{-3.5cm}
			\centering
			\includegraphics[width=\linewidth]{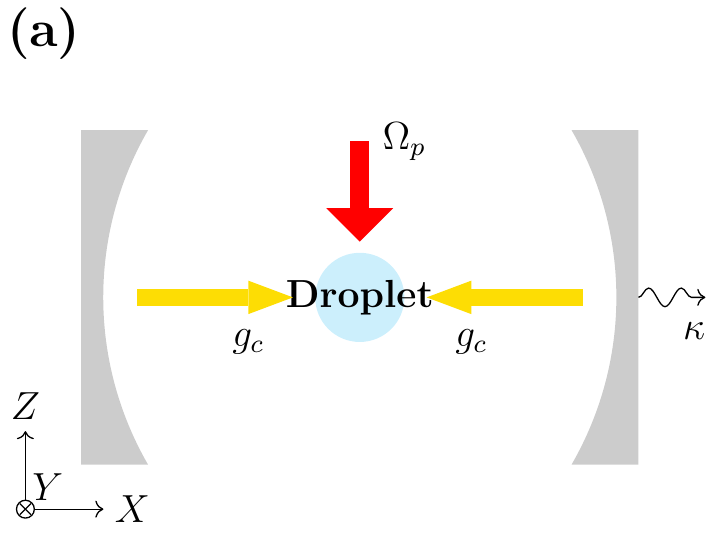}
		\end{minipage}
		\hfill
		\begin{minipage}[b]{0.48\linewidth}
			\centering
			\includegraphics[width=\linewidth]{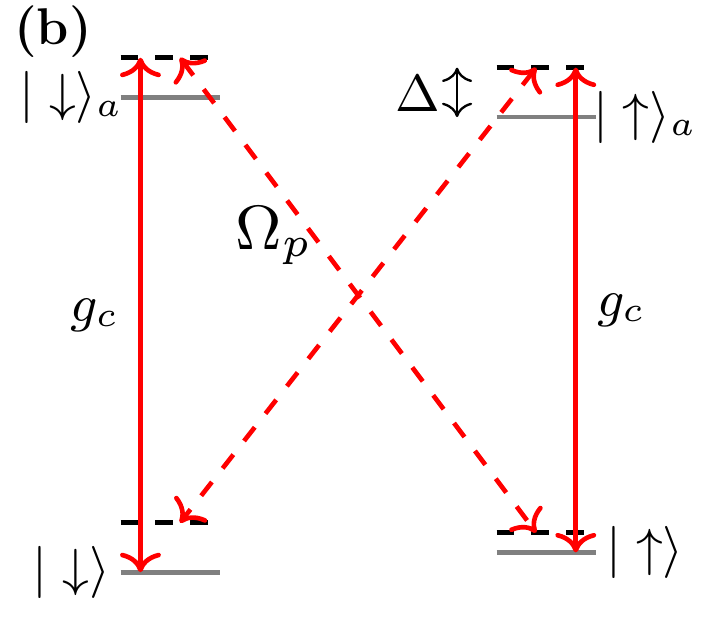}
		\end{minipage}
		
		\vspace{0.5cm}

		\begin{minipage}[b]{0.45\linewidth}
			\centering
			\includegraphics[width=\linewidth]{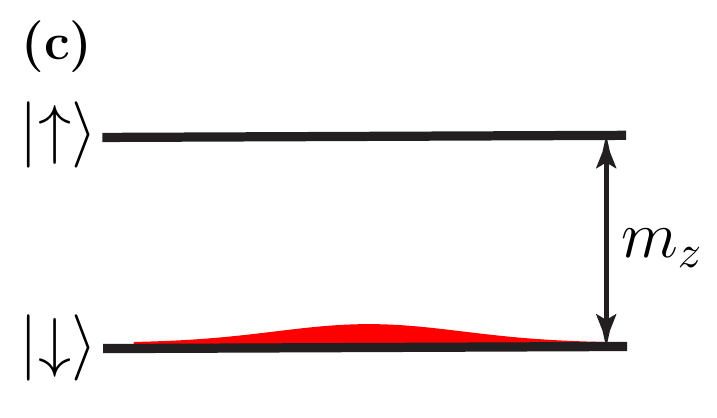}
		\end{minipage}
		\hfill
		\begin{minipage}[t]{0.5\linewidth} 
			\vspace*{-2.35cm} 
			\centering
			\includegraphics[width=\linewidth]{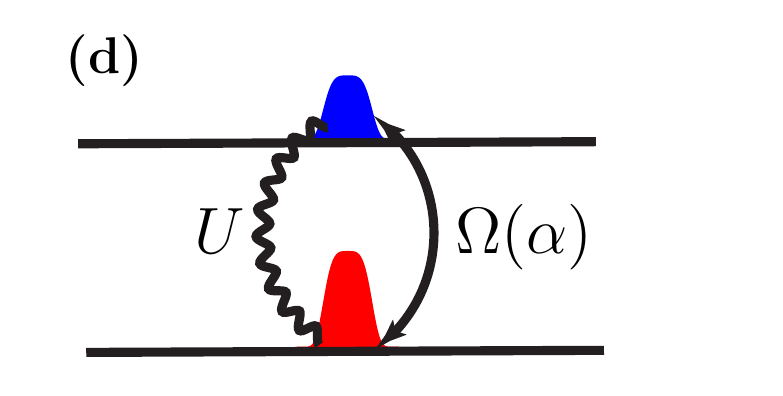}
		\end{minipage}
		
		\caption{(a) Schematic of the cavity setup. A transversely pumped microwave cavity couples
			the two components of a Bose-Einstein condensate. The pumping laser (red arrow) has a Rabi frequency $\Omega_p$, with the cavity coupling strength $g_c$ and cavity loss rate $\kappa$.
			(b) Level diagram showing the two cavity-assited Raman processes, with single-photon detuning $\Delta$.
			(c) Illustration of the fully-polarized limit, where the Zeeman field dominates, and
			all atoms occupy the $|\downarrow\rangle$ state. No cavity field is present.
			(d) Illustration of the regime with cavity-mediated droplet.
			Here the energy gain from the droplet formation is greater than the Zeeman energy bias.
			It is then favorable for a superradiant cavity to mix the two spin components, with the latter forming self-bound droplet under the interplay of mean-field interactions and quantum fluctuations.
		}
		\label{fig:fig1}
	\end{figure}

	{\it Model.---}
	As illustrated in Fig.~\ref{fig:fig1}(a), we consider a binary Bose-Einstein condensate coupled to a microwave cavity with transverse pumping. The two ground-state hyperfine spin components
	$|\uparrow\rangle$ and $|\downarrow\rangle $ are coupled through two cavity-assisted Raman processes [see Fig.~\ref{fig:fig1}(b)], mediated by metastable hyperfine states
	$|\uparrow\rangle _a $ and  $|\downarrow\rangle _a$.
	A Zeeman-energy offset $m_z$ between the spin states $|\uparrow\rangle$ and $|\downarrow\rangle$ is induced by a bias magnetic field.
	Assuming a large single-photon detuning $\Delta$ of the Raman processes, and adiabatically eliminating the intermediate states $|\uparrow\rangle _a$ and $|\downarrow\rangle _a$, we derive the effective Hamiltonian~\cite{pan2015topological}
	\begin{align}
		\hat{H} &\approx \sum_{\sigma}\int d\bm{r}\, \hat{\Psi}^{\dagger}_{\sigma}
		\biggl[\frac{\bm{p}^2}{2m} + V(\bm{r}) + \xi_c \hat{a}^{\dagger}\hat{a} + \xi_{\sigma}m_z\biggr]\hat{\Psi}_{\sigma} \notag \\
		&\quad - \Delta_c\hat{a}^{\dagger}\hat{a} + \eta\biggl[\int d\bm{r}\, \hat{\Psi}^{\dagger}_{\uparrow}
		(\hat{a} + \hat{a}^{\dagger})\hat{\Psi}_{\downarrow} + \mathrm{H.c.}\biggr] + U,
		\label{ham2}
	\end{align}
	where $\hat{\Psi}^{\dagger}_{\sigma}$ is the atomic creation operator of spin component $|\sigma\rangle$ ($\sigma\in \{\uparrow,\downarrow\}$), $\hat{a}^{\dagger}$ is the creation operator for cavity photons, $m$ is the atomic mass, $\Delta_c$ is the cavity detuning (frequency difference between the pumping and cavity fields), $\eta$ is the pumping strength,
	$\xi_{\sigma}=\pm 1$, $\xi_c=g_c^2/\Delta$, and the effective Rabi frequency of the cavity-assisted Raman processes $\eta=\Omega_p g_c/\Delta$. Here $\Omega_p$ and $g_c$ are the Rabi frequencies of the pumping and cavity fields, respectively.
	
	The interatomic interactions are given by
	\begin{align}
		U = \frac{1}{2}\int d\bm{r}\sum_{\sigma\sigma'}g_{\sigma\sigma'}\hat{\Psi}^{\dagger}_{\sigma}(\bm{r})\hat{\Psi}^{\dagger}_{\sigma'}(\bm{r})\hat{\Psi}_{\sigma'}(\bm{r})\hat{\Psi}_{\sigma}(\bm{r}),
	\end{align}
	where $g_{\sigma\sigma'}$ is the strength of contact interaction between spin species $\sigma$ and $\sigma'$, and we consider $g_{\uparrow\uparrow}, g_{\downarrow\downarrow} > 0$ and $\delta g \equiv g_{\uparrow\downarrow}+
	\sqrt{g_{\uparrow\uparrow}g_{\downarrow\downarrow}} < 0$, such that a stable quantum droplet is   supported in the absence of $g_c$, $\Omega_p$, and $m_z$, under an appropriate density ratio~\cite{petrov2015quantum}.

	Assuming a dissipative cavity with a decay rate $\kappa$, we focus on the steady-state solution of the atom-cavity hybrid system. Taking the mean-field approximation $\hat{a}\approx \langle\hat{a}\rangle:=\alpha$ and $\hat{\Psi}\approx \langle \hat{\Psi}_\sigma\rangle:=\Psi_\sigma$, the stationary condition
	$\partial\alpha/\partial t=0$ gives
	\begin{align}\label{eq:alpha}
		\alpha= \frac{\eta\left(\int d\bm{r}\, \Psi^{*}_{\uparrow}\Psi_{\downarrow} + \mathrm{H.c.}\right)}{\Delta_c + i\kappa - \xi_c N},
	\end{align}
	where $N=\sum_{\sigma}\int d \bm{r}\abs{\Psi_{\sigma}}^2$ is the total atom number.

	Substituting Eq.~(\ref{eq:alpha}) into the full Hamiltonian, and considering the Lee-Huang-Yang correction to the mean-field interaction energy, the energy functional of the system is
	\begin{align}\label{eq:efunc}
		&		E[\Psi^{*}_{\sigma},\Psi_{\sigma}]= \nonumber\\
		& \quad\sum_{\sigma}\int d\bm{r}\, \Psi^{*}_{\sigma}
		\biggl[-\frac{\hbar^2\nabla^2}{2m} + V(\bm{r}) + \xi_c \abs{\alpha}^2 - \xi_{\sigma}\delta\biggr]\Psi_{\sigma} \notag \\
		& \quad- \Delta_c\abs{\alpha}^2 - \Omega\left(\int d\bm{r}\, \Psi^{*}_{\uparrow}\Psi_{\downarrow} + \mathrm{H.c.}\right) \notag \\
		&\quad + \frac{1}{2}\int d\bm{r}\sum_{\sigma\sigma'}g_{\sigma\sigma'}\abs{\Psi_{\sigma}}^2\abs{\Psi_{\sigma'}}^2 + E_{\mathrm{LHY}}[\Psi^{*}_{\sigma},\Psi_{\sigma}],
	\end{align}
	where $\Omega= -2\eta\, \mathrm{Re}[\alpha]$, and $E_{\mathrm{LHY}}$ is the Lee-Huang-Yang energy~\cite{supp}.
	The steady-state solution of the system is then numerically obtained by minimizing the energy functional Eq.~(\ref{eq:efunc}), while self-consistently imposing the stationary condition Eq.~(\ref{eq:alpha}).

	{\it Analytical results in the homogeneous case.---}
	Let us start by examining two limiting cases for a qualitative understanding of the possible steady states. The first scenario, illustrated in Fig.~\ref{fig:fig1}(c), corresponds to the case where the Zeeman field $m_z$ dominates, so that the condensate is fully polarized in the low-energy spin state $|\downarrow\rangle$. According to Eq.~(\ref{eq:alpha}), no cavity field is generated in this case, and the condensate
	is in the gas phase.
	The second scenario, illustrated in Fig.~\ref{fig:fig1}(d), involves a sufficiently large pumping $\eta$, such that the system becomes superradiant with a finite cavity field $\alpha$.
	The two spin components are then mixed under the finite cavity-assisted Raman coupling $\Omega$.
	The inter-species attraction $g_{\uparrow\downarrow}$, combined with quantum fluctuations (in the form of the Lee-Huang-Yang correction), can stabilize a self-bound quantum droplet.
	Thus, in between these two limits, a gas-liquid transition should exist.

	We now analyze the energy functional Eq.~(\ref{eq:efunc}) in a homogeneous setting.
	In the gas phase, we label the density ratio between the two spin components as $R :=n_\uparrow / n_\downarrow$. The densities of the two spin species are then $n^g_\uparrow = n^gR/(1+R)$ and $n^g_\downarrow = n^g/(1+R)$, where $n^g$ is the total number density.
	From Eq.~(\ref{eq:alpha}), the steady-state cavity field is
	\begin{equation}
		\alpha \approx \frac{-2\sqrt{R}}{\xi_c(1 + R)} \eta,
		\label{alp1}
	\end{equation}
	and the average energy per particle is
	\begin{align}
		\frac{E^g(R)}{N} &=  \frac{E^g_{\mathrm{LHY}}(R)}{N}-\frac{4\eta^2 R }{\xi_c(1+R)^2}
		+ \frac{m_z (R - 1)}{1 + R} \notag\\
		&\quad + \left( \frac{1}{2} g_{\uparrow\uparrow} R^2 + g_{\uparrow\downarrow} R + \frac{1}{2} g_{\downarrow\downarrow} \right) \frac{n^g}{(1 + R)^2}
		,
		\label{Eg}
	\end{align}
	where $E^g_{\mathrm{LHY}}(R)$ denotes the Lee-Huang-Yang correction in the gas phase. The steady-state solution can be solved by minimizing $E^g(R)$ with respect to $R$.
	
	In the liquid phase, the quantum droplet has a fixed density ratio \( R_P = \sqrt{g_{\downarrow\downarrow} / g_{\uparrow\uparrow}} \), but a variable density $n^d$.
	It follows that the energy functional in the liquid phase is obtained by substituting $n^g$ and $R$ in Eq.~(\ref{Eg}) with $n^d$ and $R_P$, respectively. The steady state is solved by minimizing the resulting $E^d(n^d)$ with respect to $n^d$.

	\begin{figure}[tbp]
		\centering
		\hspace*{-1.1cm}
		\begin{minipage}[b]{0.494\linewidth}
			\centering
			\includegraphics[width=1.15\linewidth]{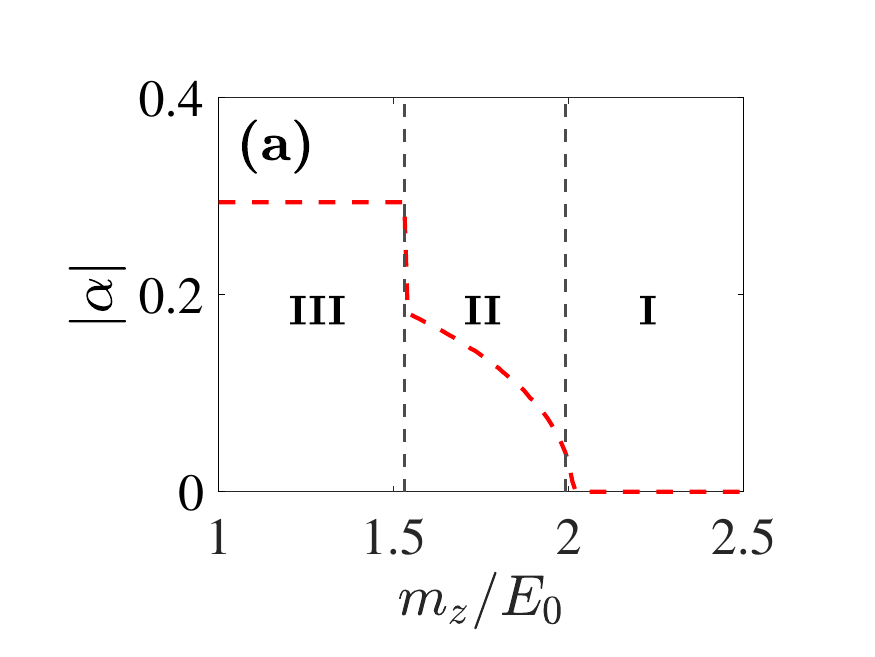}
		\end{minipage}
		\begin{minipage}[b]{0.494\linewidth}
			\centering
			\includegraphics[width=1.15\linewidth]{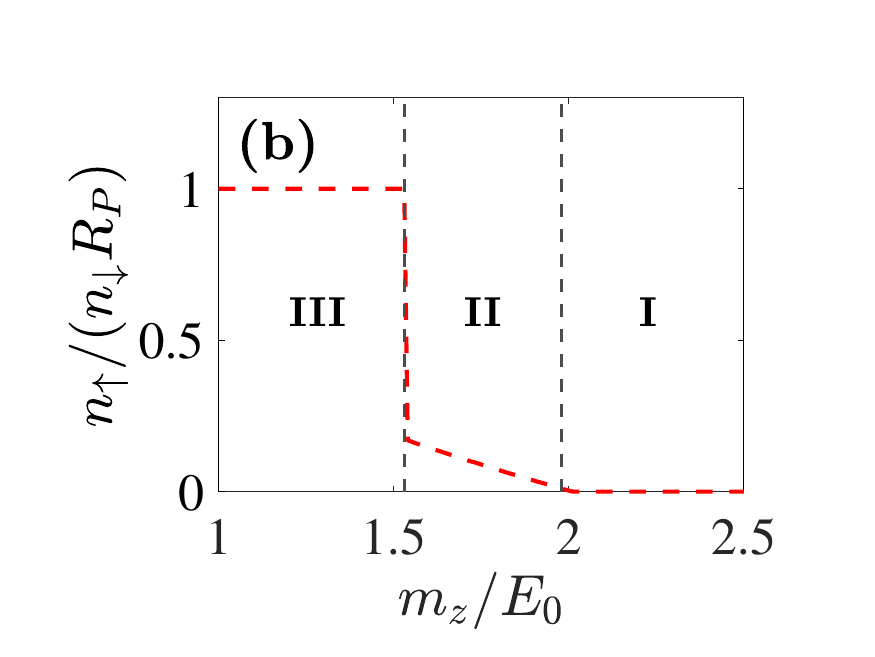}
		\end{minipage}
		
		\hspace*{-1.1cm}
		\begin{minipage}[b]{0.494\linewidth}
			\centering
			\includegraphics[width=1.15\linewidth]{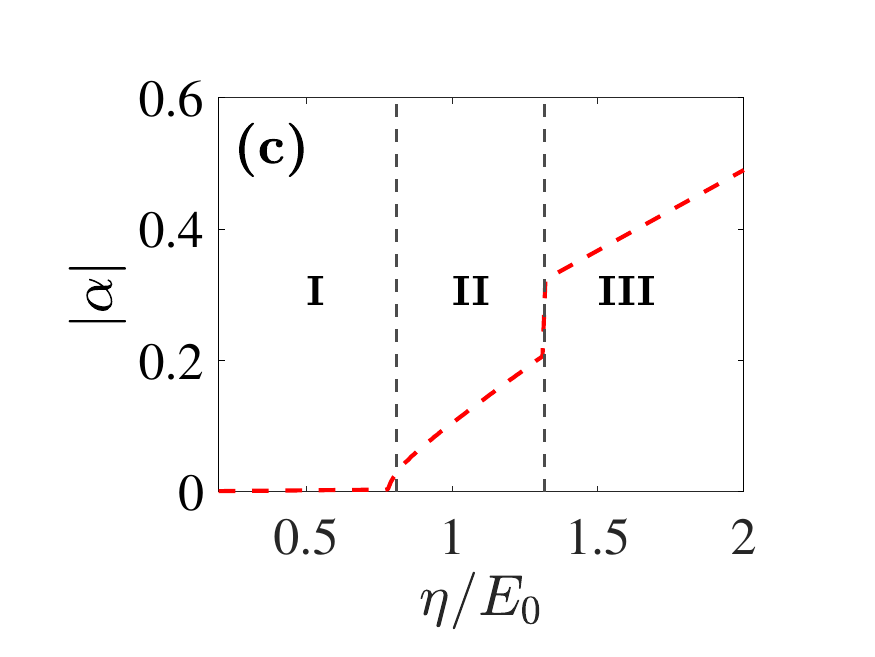}
		\end{minipage}
		\begin{minipage}[b]{0.494\linewidth}
			\centering
			\includegraphics[width=1.15\linewidth]{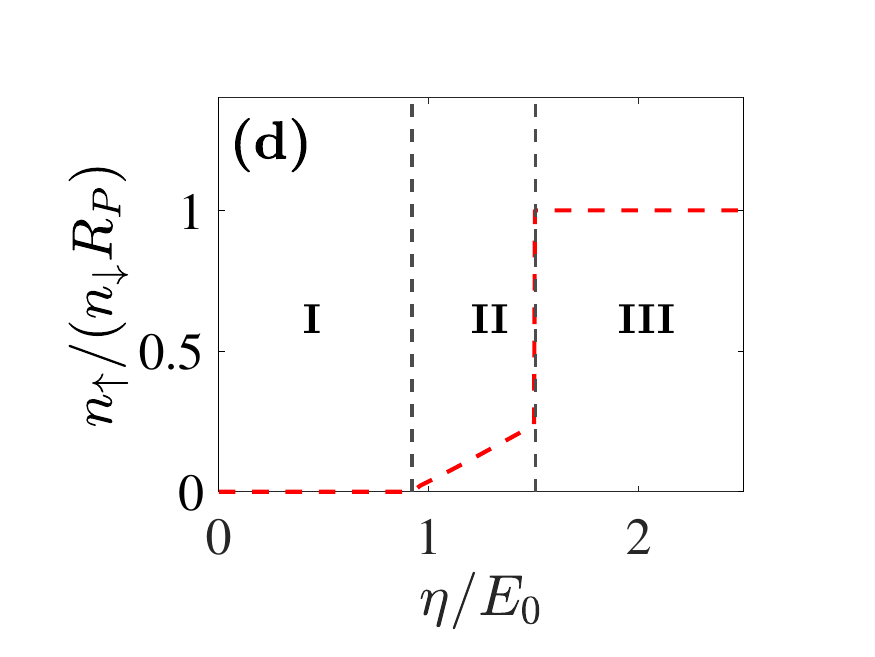}
		\end{minipage}
		
		
		\caption{(a)(b) Cavity field \( |\alpha| \) and density ratio \( n_{\uparrow}/n_{\downarrow} \) as functions of \( m_z \) at a fixed \( \eta = 1.2\,E_0 \). The superradiant transition occurs at \( m_z = 1.99\,E_0 \), followed by a gas-liquid transition at \( m_z = 1.53\,E_0 \). (c)(d) Cavity field \( |\alpha| \) and density ratio \( n_{\uparrow}/n_{\downarrow} \) as functions of \( \eta \) at a fixed \( m_z = 1.6\,E_0 \). The superradiant transition occurs at \( \eta_c = 0.81\,E_0 \), followed by the gas-liquid transition at \( \eta= 1.32\,E_0 \). In all panels, we set \( \xi_c = 4\,E_0 \). The unit of energy is defined through the mean-field interaction energy $E_0:=n^g g_{\downarrow\downarrow}$.}
		\label{fig:fig2}
	\end{figure}

	In Fig.~\ref{fig:fig2}, we show the steady-state cavity field [see Fig.~\ref{fig:fig2}(a)(c)] and density ratio [see Fig.~\ref{fig:fig2}(b)(d)], as functions of the Zeeman field $m_z$ and pumping strength, respectively.
	Here three different phase regimes are visible. When $m_z$ is sufficiently large, the cavity is not superradiant and the condensate is in a spin polarized gas phase (phase I). When $m_z$ is small, the cavity is superradiant and a droplet with fixed density ratio emerges (phase III). These are the two limiting cases that we expect. In between the two, a third phase (phase II) arises, with a superradiant cavity, and a partially spin polarized gas phase in the condensate.
	Importantly, between phase III and phase II, the superradiant cavity field undergoes an abrupt jump at the phase boundary [see Fig.~\ref{fig:fig2}(a)]. Further, in phase III, the cavity field scales linearly with the pumping strength [see Fig.~\ref{fig:fig2}(c)]. Both of these features originate from the stationary condition Eq.~(\ref{alp1}). Whereas the abrupt jump in the cavity field reflects the first-order nature of the transition with similar abrupt jumps in the density ratio [see Fig.~~\ref{fig:fig2}(b)(d)], the linear scaling derives from the fixed density ratio of quantum droplet.

	On the other hand, it is evident that the critical pumping strength of the superradiant transition $\eta_c$ is dependent on the Zeeman field.
	In fact, below a critical Zeeman field $m^c_z$, the superradiance is dramatically enhanced thanks to the simultaneous droplet formation, such that it occurs at an infinitesimally small pumping strength. For a qualitative understanding, let us consider the small-pumping limit $\eta\rightarrow 0$. Above $m^c_z$, the Zeeman field is so strong that the system should energetically favor a spin-polarized steady state (phase I or phase II).
	However, below $m^c_z$, the energy gain by forming a quantum droplet can outweigh the Zeeman-energy cost of spin mixing (mediated by cavity-assisted Raman couplings).
	But according to Eq.~(\ref{alp1}), the droplet formation is predicated on the presence of cavity field. Hence, the system becomes superradiant with the simultaneous formation of cavity-meidated quantum droplet (phase III), even under an arbitrarily small $\eta$.

	According to the analysis above, the critical Zeeman field $m_z^c$ can be solved by equating the Zeeman energy offset with the energy gain from the droplet formation. Specifically, in the spin-fully-polarized gas phase, we have \( R = 0 \), under which the average energy per particle becomes
	\begin{equation}
		\frac{E^g}{N} \approx -m_z+ \frac{1}{2} g_{\downarrow\downarrow} n^g. \label{Ega}
	\end{equation}
	In the droplet phase and in the limit \( \eta \to 0 \),
	the mean-field energy is
	\begin{align}
		\frac{E^d_{\mathrm{MF}}(n^d)}{N}&=\left( \frac{1}{2} g_{\uparrow\uparrow} R_P^2 + g_{\uparrow\downarrow} R_P + \frac{1}{2} g_{\downarrow\downarrow} \right) \frac{n^d}{(1 + R_P)^2}\nonumber\\
		&+ \frac{m_z (R_P - 1)}{1 + R_P}\nonumber\\
		&:=\frac{E^{d0}_{\mathrm{MF}}(n^d)}{N}+ \frac{m_z (R_P - 1)}{1 + R_P},
	\end{align}
	with the Lee-Huang-Yang correction~\cite{petrov2015quantum}
	\begin{align}
		\frac{E^d_{\mathrm{LHY}}(n^d)}{N}& = \frac{\sqrt{2m^3}}{15\pi^2\hbar^3n^d}\Big[g_{\uparrow\uparrow}n^d_\uparrow+g_{\downarrow\downarrow}n^d_\downarrow\notag\\&+ \sqrt{(g_{\uparrow\uparrow}n^d_\uparrow-g_{\downarrow\downarrow}n^d_\downarrow)^2+4g_{\uparrow\downarrow}^2 n^d_\uparrow n^d_\downarrow}\,\Big]^{5/2}.
	\end{align}
	Thus, the critical Zeeman field is
	\begin{equation}
		m_z^c=\left(\frac{R_P+1}{2R_P}\right)\left(\frac{1}{2} g_{\downarrow\downarrow} n^g-\frac{E^d_{\mathrm{min}}}{N}\right),\label{mzc}
	\end{equation}
	where $E^d_{\mathrm{min}}$ denotes the minimum of \( E^{d0}_{\mathrm{MF}}(n^d) + E^d_{\mathrm{LHY}}(n^d) \).

	\begin{figure}[tbp]
		\centering
		\hspace*{-0.12\linewidth}  
		\begin{minipage}[b]{1.23\linewidth}  
			\centering
			\includegraphics[width=0.8\linewidth]{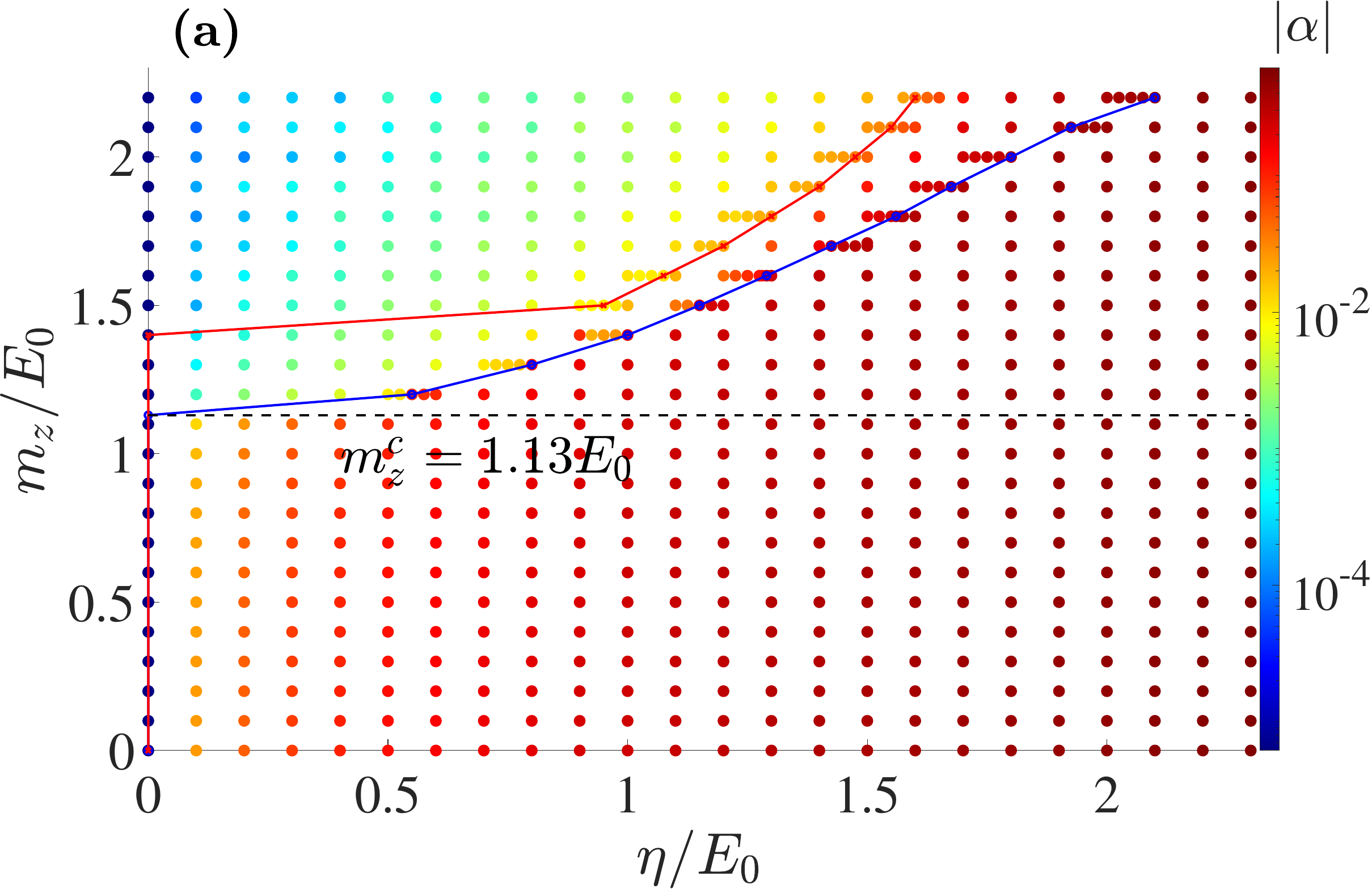}
		\end{minipage}
		
		\vspace{1em}  
		
		\begin{minipage}[b]{0.48\linewidth}
			\centering
			\includegraphics[width=\linewidth]{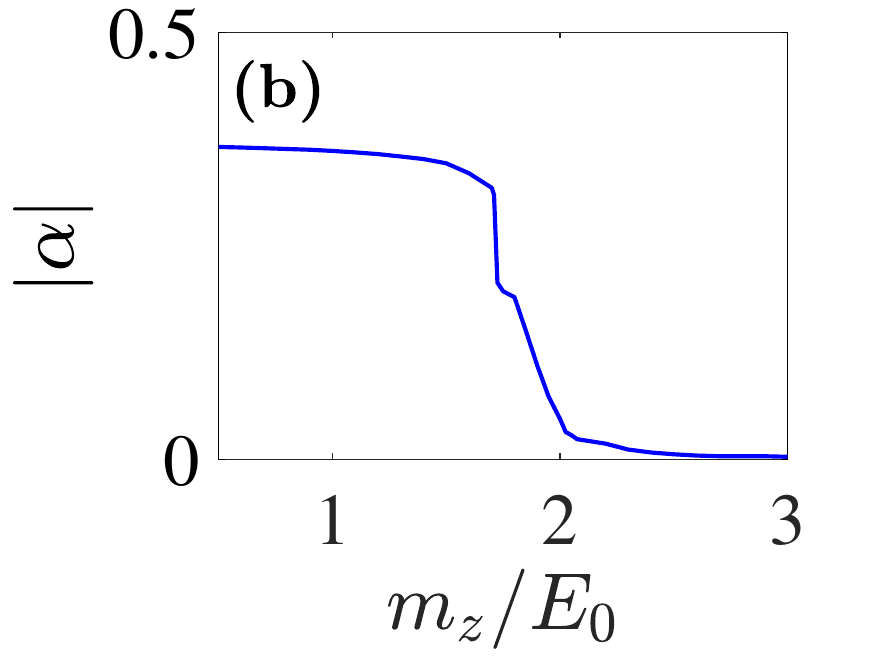}
		\end{minipage}
		\hfill
		\begin{minipage}[b]{0.48\linewidth}
			\centering
			\includegraphics[width=\linewidth]{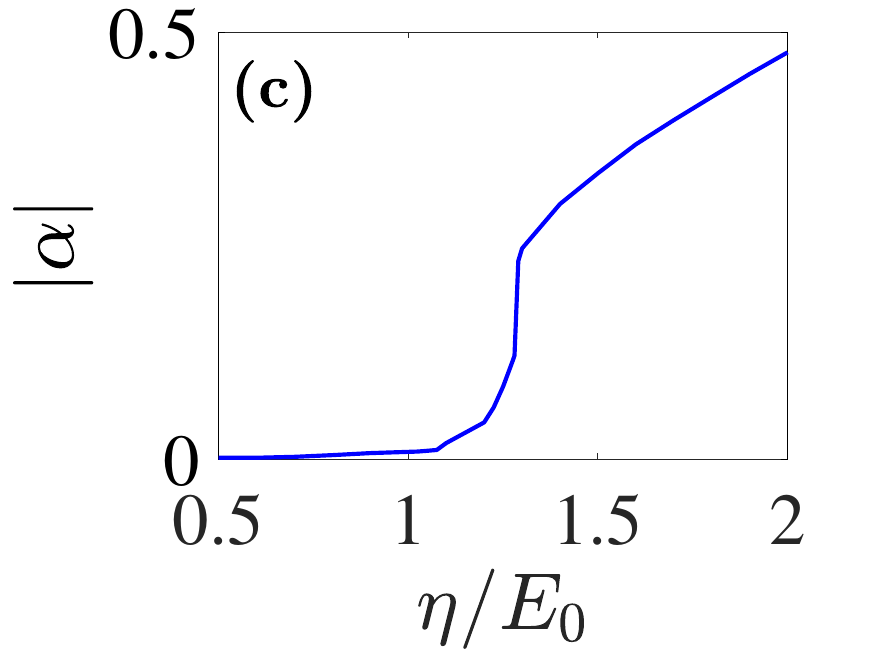}
		\end{minipage}
		
		\caption{(a) Thermal map of the steady-state cavity field $|\alpha|$ as a function of $m_z$ and $\eta$. The red line indicates the superradiant phase transition, while the blue line marks the gas-liquid transition. The intercept of the blue line is $m_z^c = 1.13 E_0$.
			(b) Cavity field $|\alpha|$ as a function of $m_z$ with a fixed $\eta = 1.5 E_0$. The superradiant transition occurs at $m_z = 2.03 E_0$, and the gas-liquid transition is at $m_z = 1.71 E_0$.
			(c) Cavity field $|\alpha|$ as a function of $\eta$ with a fixed $m_z = 1.6 E_0$. The superradiant and gas-liquid transitions are located at $\eta = 1.08 E_0$ and $\eta = 1.29 E_0$, respectively.}
		\label{fig:fig3}
	\end{figure}
	For an estimate of $m^c_z$, we take the typical parameters of $^{39}$K, where the spin states $|\uparrow\rangle$
	and $|\downarrow\rangle$ correspond to the ground hyperfine states $|F=1,m_F=0\rangle$ and $F=1,m_F=1\rangle$, respectively. The interatomic contact interactions are characterized by the \( s \)-wave scattering lengths: \( a_{\uparrow\uparrow} = 74.9834\,a_0 \), \( a_{\downarrow\downarrow} = 33.5\,a_0 \), and \( a_{\uparrow\downarrow} = -53.1418\,a_0 \), where \( a_0 \) is the Bohr radius. The corresponding interaction strengths are given by \( g_{\sigma\sigma'} = 4\pi\hbar^2 a_{\sigma\sigma'}/m \).
	Fixing the total number density at \( n^g = 1.3 \times 10^{19}~\mathrm{m}^{-3} \), we have \( m_z^c = 1.14\,E_0 \), where we take the unit of energy to be the interaction energy $E_0 = n^g g_{\downarrow\downarrow} = 75.13\,\mathrm{Hz}$. As such, the presence of a finite $m^c_z$ confirms the enhanced superradiance, which is underlain by the interplay of the Zeeman field, the cavity-assisted Raman coupling, and droplet formation.
	
	{\it Phase diagram within a harmonic trap.---}
	We now confirm the analytical results above with numerical simulations. For experimental relevance, we consider an external harmonic trapping potential of the form $V(\bm{r}) = \frac{1}{2} m \omega^2 r^2$ for the condensate. The trapping frequency satisfies $\omega \gg \hbar / (m \lambda_0^2)$, so that the characteristic length of the trapping potential is much smaller than the wavelength of the cavity field $\lambda_0$ and we can neglect the spatial variation of the cavity field across the expanse of the condensate. In the presence of the trap, the Lee-Huang-Yang (LHY) correction is included under the local density approximation~\cite{supp}.

	For numerical calculations, we take typical parameters of $^{39}\mathrm{K}$ atoms, with $N = 2.95 \times 10^5$, trapping frequency $\omega = 2\pi \times  7.96\,\mathrm{Hz}$, cavity detuning $\Delta_c = -80 E_0$, cavity coupling $\xi_c = 4 E_0$, cavity photon loss rate $\kappa = 4 \times 10^3 E_0$, and single-photon detuning from intermediate states $\Delta = 4 \times 10^4 E_0$.

	We map out the steady-state phased diagram in Fig~\ref{fig:fig3}(a), showing the thermal map of the steady-state cavity field $|\alpha|$ under discrete sets of parameters $(m_z,\eta)$.
	Two phase boundaries are identified: the red solid line marking the superradiant phase transition where the cavity field becomes finite, and the blue solid line indicating the gas-liquid phase transition where a self-bound droplet emerges.
	Consistent with the analysis in the homogeneous case, the phase diagram is separated into two distinct regions by a critical Zeeman field $m^c_z\approx 1.13 E_0$ (close to the previously estimated value $1.14 E_0$).
	When $m_z < m_z^c$, superradiance is enhanced, as an infinitesimally small pumping strength $\eta$ can stabilize phase III, leading simultaneously to superradiance and droplet formation.
	By contrast, when $m_z>m_z^c$, the Zeeman field strongly favors a polarized gas (phase I or phase II), and droplet formation only takes place after the onset of superradiance. 
	Notably, under intermediate $m_z$, superradiance is still enhanced, where an infinitesimally small pumping strength would drive the system into a partially polarized gas with superradiant cavity (phase II).

	In Figs.~\ref{fig:fig3}(b)(c), we show the variations of the cavity field $|\alpha|$ with increasing Zeeman field and pumping strength, respectively.
	Importantly, key signals of the cavity field persist in the presence of trapping potential:
	the cavity field undergoes a sharp jump at the gas-liquid transition, and features a linear scaling with the pumping strength in the liquid phase.

	\begin{figure}[tbp]
		\centering
		\hspace*{-0.9cm}
		\begin{minipage}[t]{0.494\linewidth}
			\centering
			\includegraphics[width=1.1\linewidth]{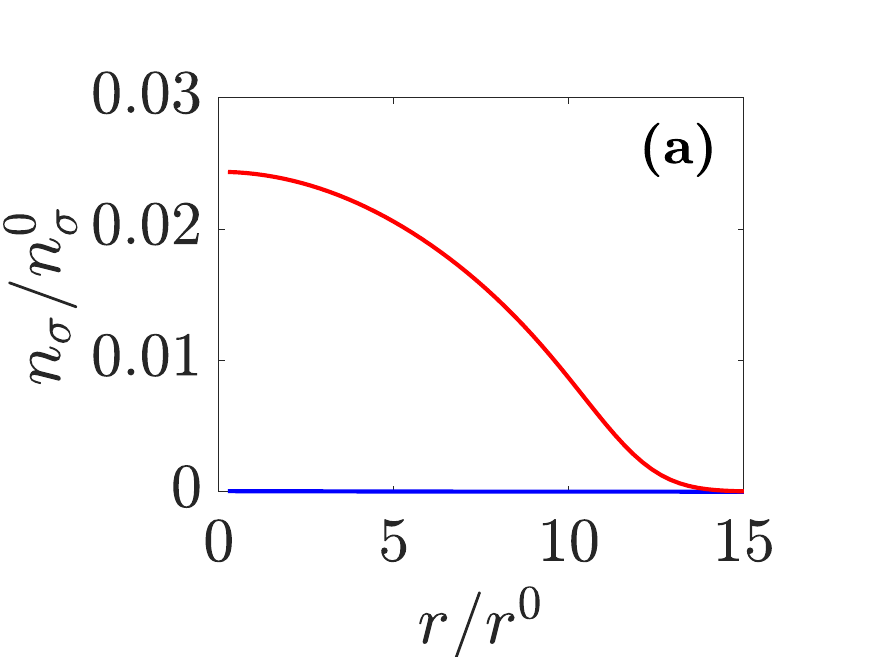}
		\end{minipage}
		\begin{minipage}[b]{0.494\linewidth}
			\centering
			\includegraphics[width=1.1\linewidth]{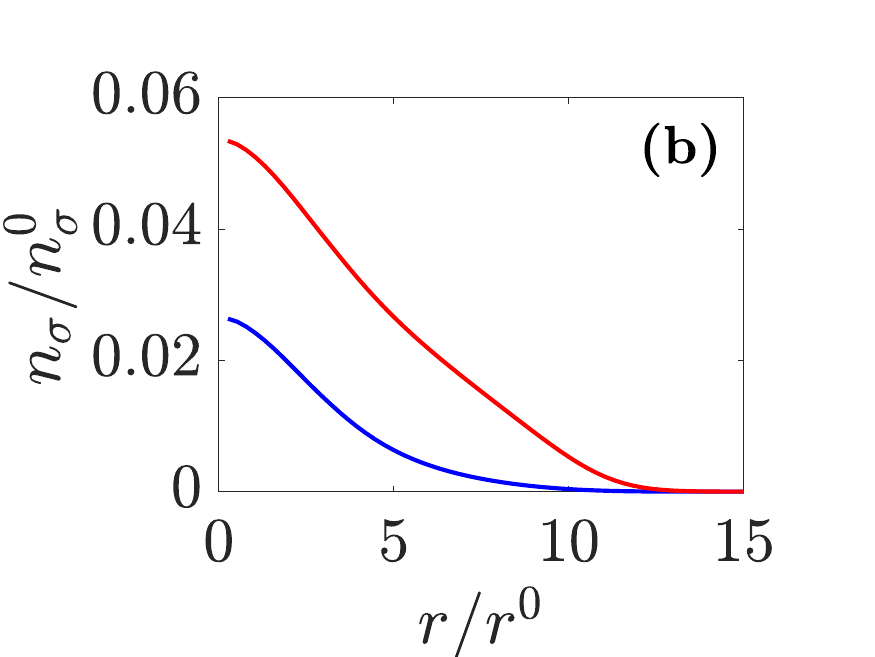}
		\end{minipage}
		\hspace*{-0.9cm}
		\begin{minipage}[b]{0.494\linewidth}
			\centering
			\includegraphics[width=1.1\linewidth]{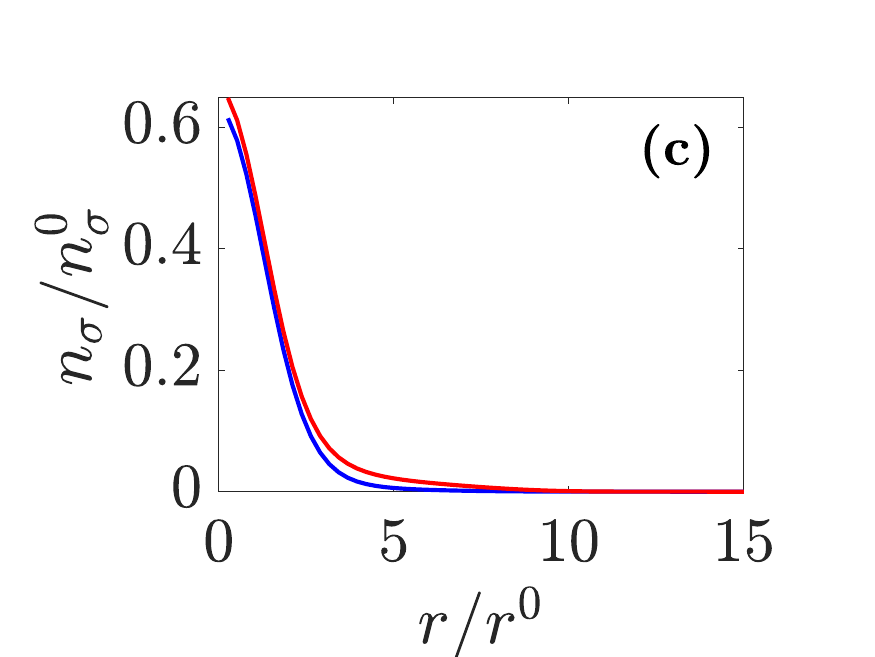}
		\end{minipage}
		\begin{minipage}[t]{0.494\linewidth}  
			\centering
			\includegraphics[width=1.1\linewidth]{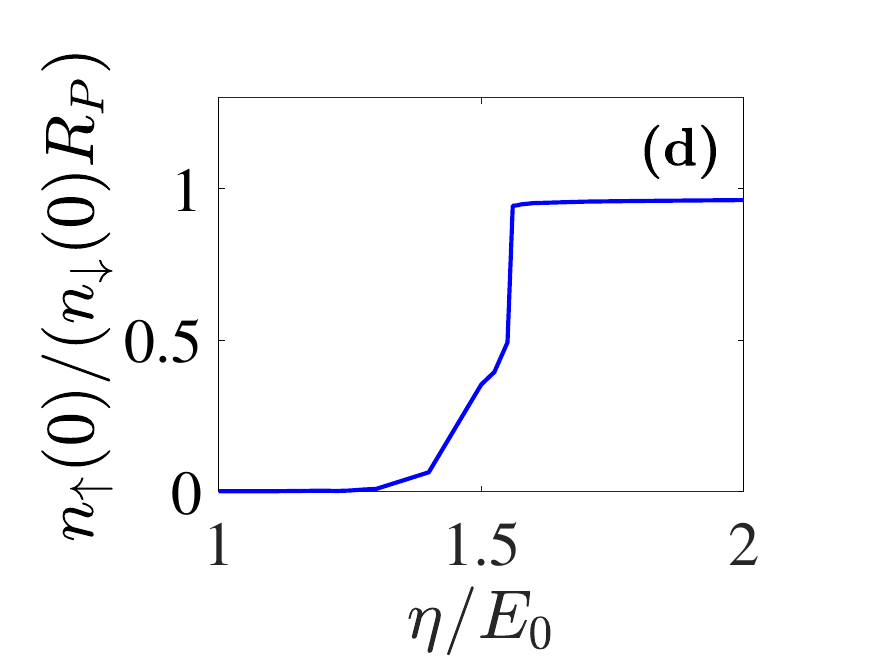}
		\end{minipage}
		
		\caption{ 
			Real-space density profiles of the two spin components in a trapping potential.
			(a) Phase I (fully polarized gas) at $\eta = 1 E_0$ and $m_z = 1.8 E_0$.
			(b) Phase II (partially polarized gas with superradiance) at $\eta = 1.55 E_0$ and $m_z = 1.8 E_0$.
			(c) Phase III (self-bound droplet with superradiance) at $\eta = 1.56 E_0$ and $m_z = 1.8 E_0$.
			(d) Central density ratio $n_{\uparrow}(r=0)/n_{\downarrow}(r=0)$ as functions of increasing $\eta$ with a fixed $m_z = 1.8 E_0$.
			Here, we define the reference densities as $n_{\uparrow}^0 = 26.57n^g$ and $n_{\downarrow}^0 = n_{\uparrow}^0 / R_P$.
		}
		
		\label{fig:fig4}
	\end{figure}
	
	Finally, we plot the spin-resolved density profiles of different phases in Fig.~\ref{fig:fig4}(a)(b)(c). Particularly, in phase III [see Fig.~\ref{fig:fig4}(c)], the droplet core is surrounded by a polarized shell of gaseous condensate, suggesting gas-liquid coexistence.
	However, it is generally difficult to identify the onset of the droplet core from the density profiles alone.
	While the phase transitions can in principle be revealed through the spin density ratio at the trap center [see Fig.~\ref{fig:fig4}(d)], the measurement requires \textit{in situ} imaging and is challenging.
	Instead, the abrupt jump and the linear scaling of the cavity field can serve as prominent signals for the gas-liquid transition and coexistence.

	{\it Discussion.---}
	We show that, by imposing cavity-assisted Raman coupling between the two hyperfine components
	of a binary Bose-Einstein condensate, the gas-liquid transition and coexistence leave clear signals in the cavity field, offering convenient detection schemes.
	We also reveal an enhanced superradiance transition in our configuration, which originates from the	interplay of Zeeman field, cavity-assisted coupling, and quantum fluctuations in the atom sector.
	This is complementary to previous studies on quantum-droplet formation through cavity-induced long-range interactions, where cavity fluctuations play a key role~\cite{masalaeva2023tuning,mixa2025engineering}.
	In the context of recent studies on quantum criticality in the gas-liquid transition of binary Bose-Einstein condensate~\cite{gu2023liquid,he2023quantum},
	it is tempting to further explore the critical behaviors herein, which are enriched by the action and back action between the matter and cavity fields.
	It is also intriguing to investigate the coupling of droplet arrays with the cavity, where quantum fluctuations of both the cavity field and atoms can lead to exotic quantum matter.

	\begin{acknowledgments}
		We thank Qi Gu for valuable discussions. This work is supported by the National Natural Science Foundation of China (Grant No. 12374479), and by the Innovation Program for Quantum Science and Technology (Grant No. 2021ZD0301904).
	\end{acknowledgments}


	\bibliographystyle{apsrev4-2}
	\bibliography{Ref1}
	
	\clearpage
	\onecolumngrid
	
	\vspace*{1cm}
	\begin{center}
		\textbf{\Large Supplemental Material}
	\end{center}
	
	\vspace{0.5cm}
	
	\noindent
	In this Supplemental Material, we provide details on the derivation of the Lee-Huang-Yang corrections in the presence of Raman coupling \(\Omega\) and Zeeman field \(\delta\).
	
	\section*{Lee-Huang-Yang Correction: Bogoliubov Approach with Raman Coupling}
	
	We here outline the derivation of the beyond-mean-field Lee-Huang-Yang correction in the presence of both the cavity-assisted Raman coupling \(\Omega\) and Zeeman detuning \(\delta\).
	
	For a homogeneous two-component Bose gas, the system Hamiltonian in momentum space reads \( H = H_0 + U \), where \(H_0\) contains the kinetic, Raman, and Zeeman terms, and \(U\) accounts for the contact interactions
	\begin{align}
		H_0 = \sum_{\mathbf{k}} \Big[
		\epsilon_{\mathbf{k}} \left( \psi_{\uparrow,\mathbf{k}}^\dagger \psi_{\uparrow,\mathbf{k}} + \psi_{\downarrow,\mathbf{k}}^\dagger \psi_{\downarrow,\mathbf{k}} \right)
		- \Omega \left( \psi_{\uparrow,\mathbf{k}}^\dagger \psi_{\downarrow,\mathbf{k}} + \psi_{\downarrow,\mathbf{k}}^\dagger \psi_{\uparrow,\mathbf{k}} \right)
		- \delta \left( \psi_{\uparrow,\mathbf{k}}^\dagger \psi_{\uparrow,\mathbf{k}} - \psi_{\downarrow,\mathbf{k}}^\dagger \psi_{\downarrow,\mathbf{k}} \right) \notag
		\Big],
	\end{align}
	
	\begin{align}
		U = \frac{1}{2V} \sum_{\mathbf{q},\mathbf{p},\mathbf{k}} \sum_{\sigma\sigma'}
		g_{\sigma\sigma'} \psi_{\sigma,\mathbf{q}+\mathbf{k}}^\dagger \psi_{\sigma',\mathbf{q}-\mathbf{k}}^\dagger
		\psi_{\sigma',\mathbf{q}-\mathbf{p}} \psi_{\sigma,\mathbf{q}+\mathbf{p}}. \tag{S1}
	\end{align}
	Here $\psi_{\sigma,\mathbf{k}}$ is the annihilation operator for spin species $\sigma$ and momentum $\mathbf{k}$, and $V$ is the quantization volume.
	
	Following the standard Bogoliubov prescription, we perturb around the mean-field ground state and retain quadratic terms in the fluctuation. The quadratic Hamiltonian becomes
	\begin{align}
		\frac{H}{V} = \frac{E_{\mathrm{MF}}}{V} + \frac{1}{2V} \sum_{\sigma\sigma'} g^2_{\sigma\sigma'} n_\sigma n_{\sigma'}\sum_{\mathbf{k} \ne 0}\frac{1}{2\epsilon_{\mathbf{k}}}
		- \frac{1}{2V}\sum_{\mathbf{k} \ne 0} 
		\left(
		2\epsilon_{\mathbf{k}} + \Omega \frac{(n_\uparrow + n_\downarrow)}{\sqrt{n_\uparrow n_\downarrow}} + g_{\uparrow\uparrow} n_\uparrow + g_{\downarrow\downarrow} n_\downarrow
		\right)
		+ \frac{1}{2V} \sum_{\mathbf{k} \ne 0} A^\dagger H_{\mathrm{Bog}} A, \tag{S2}
	\end{align}
	where \( A^\dagger = (\psi_{\uparrow,\mathbf{k}}^\dagger, \psi_{\uparrow,-\mathbf{k}}, \psi_{\downarrow,\mathbf{k}}^\dagger, \psi_{\downarrow,-\mathbf{k}}) \).
	
	The Bogoliubov matrix \(H_{\mathrm{Bog}}\) takes the form
	\begin{align}
		H_{\mathrm{Bog}} =
		\begin{pmatrix}
			\epsilon_{\mathbf{k}} + \Omega \sqrt{\frac{n_\downarrow}{n_\uparrow}} + g_{\uparrow\uparrow} n_\uparrow &
			g_{\uparrow\uparrow} n_\uparrow &
			g_{\uparrow\downarrow} \sqrt{n_\uparrow n_\downarrow} - \Omega &
			g_{\uparrow\downarrow} \sqrt{n_\uparrow n_\downarrow} \\
			g_{\uparrow\uparrow} n_\uparrow &
			\epsilon_{\mathbf{k}} + \Omega \sqrt{\frac{n_\downarrow}{n_\uparrow}} + g_{\uparrow\uparrow} n_\uparrow &
			g_{\uparrow\downarrow} \sqrt{n_\uparrow n_\downarrow} &
			g_{\uparrow\downarrow} \sqrt{n_\uparrow n_\downarrow} - \Omega \\
			g_{\uparrow\downarrow} \sqrt{n_\uparrow n_\downarrow} - \Omega &
			g_{\uparrow\downarrow} \sqrt{n_\uparrow n_\downarrow} &
			\epsilon_{\mathbf{k}} + \Omega \sqrt{\frac{n_\uparrow}{n_\downarrow}} + g_{\downarrow\downarrow} n_\downarrow &
			g_{\downarrow\downarrow} n_\downarrow \\
			g_{\uparrow\downarrow} \sqrt{n_\uparrow n_\downarrow} &
			g_{\uparrow\downarrow} \sqrt{n_\uparrow n_\downarrow} - \Omega &
			g_{\downarrow\downarrow} n_\downarrow &
			\epsilon_{\mathbf{k}} + \Omega \sqrt{\frac{n_\uparrow}{n_\downarrow}} + g_{\downarrow\downarrow} n_\downarrow
		\end{pmatrix}. \tag{S3}
	\end{align}
	
	This is diagonalized using the Bogoliubov transformation, yielding
	\begin{align}
		\frac{1}{2} A^\dagger H_{\mathrm{Bog}} A
		= \frac{1}{2}  \left[ \mathcal{E}_{+,\mathbf{k}} \left( b_{+,\mathbf{k}}^\dagger b_{+,\mathbf{k}} + b_{+,-\mathbf{k}} b_{+,-\mathbf{k}}^\dagger \right)
		+ \mathcal{E}_{-,\mathbf{k}} \left( b_{-,\mathbf{k}}^\dagger b_{-,\mathbf{k}} + b_{-,-\mathbf{k}}b_{-,-\mathbf{k}}^\dagger  \right)
		\right] , \tag{S4}
	\end{align}
	where \( b_{\pm,\mathbf{k}} \) are the Bogoliubov quasiparticle operators, and the excitation energies \(\mathcal{E}_{\pm,\mathbf{k}}\) satisfy
	\begin{align}
		\det\left| H_{\mathrm{Bog}} - \mathcal{E} \begin{pmatrix} \sigma_z & 0 \\ 0 & \sigma_z \end{pmatrix} \right| = 0. \tag{S5}
	\end{align}
	
	The analytic form of the excitation energies is
	\begin{align}
		\mathcal{E}_{\pm,\mathbf{k}} = \sqrt{ D_{\mathbf{k}}^2 \pm \left\{D_{\mathbf{k}}^2 - \epsilon_{\mathbf{k}}\left( \epsilon_{\mathbf{k}} + \Omega \frac{n_\uparrow + n_\downarrow}{\sqrt{n_\uparrow n_\downarrow}} \right)\left[\left( \epsilon_{\mathbf{k}} + 2g_{\uparrow\uparrow} n_\uparrow + \Omega \sqrt{\frac{n_\downarrow}{n_\uparrow}} \right)
			\left( \epsilon_{\mathbf{k}} + 2g_{\downarrow\downarrow} n_\downarrow + \Omega \sqrt{\frac{n_\uparrow}{n_\downarrow}} \right)
			- \left( 2g_{\uparrow\downarrow} \sqrt{n_\uparrow n_\downarrow} - \Omega \right)^2\right]	\right\}^{1/2} }, \tag{S6}
	\end{align}
	where
	\begin{align}
		D_{\mathbf{k}} = \frac{1}{2} \left( \epsilon_{\mathbf{k}} + g_{\uparrow\uparrow} n_\uparrow + \Omega \sqrt{\frac{n_\downarrow}{n_\uparrow}} \right)^2
		+ \frac{1}{2} \left( \epsilon_{\mathbf{k}} + g_{\downarrow\downarrow} n_\downarrow + \Omega \sqrt{\frac{n_\uparrow}{n_\downarrow}} \right)^2+\left(g_{\uparrow\downarrow}\sqrt{n_{\uparrow}n_{\downarrow}}-\Omega\right)^2
		- \frac{1}{2} \sum_{\sigma\sigma'} g_{\sigma\sigma'}^2 n_\sigma n_{\sigma'}. \tag{S7}
	\end{align}
	
	Finally, the total energy density including quantum fluctuations is given by
	\begin{align}
		\frac{E}{V} &= \frac{E_{\mathrm{MF}}}{V}
		+ \frac{1}{2V}\sum_{\mathbf{k} \ne 0} \left[\sum_{\sigma\sigma'} g^2_{\sigma\sigma'} n_\sigma n_{\sigma'}\frac{1}{2\epsilon_{\mathbf{k}}}
		-\left( 2\epsilon_{\mathbf{k}} + \Omega \frac{(n_\uparrow + n_\downarrow)}{\sqrt{n_\uparrow n_\downarrow}} + g_{\uparrow\uparrow} n_\uparrow + g_{\downarrow\downarrow} n_\downarrow \right)
		+  \mathcal{E}_{+,\mathbf{k}} + \mathcal{E}_{-,\mathbf{k}} 
		\right]\notag\\
		&:=\frac{E_{\mathrm{MF}}}{V}+\varepsilon_{\mathrm{LHY}}.\tag{S8}\label{S8}
	\end{align}
	This expression provides the Lee-Huang-Yang correction that explicitly incorporates the effects of both the cavity-assisted Raman coupling \(\Omega\) and the Zeeman shift \(\delta\).
	
	In the presence of an external trapping potential $V(\bm{r})$, the local density at position $\bm{r}$ is denoted by $n_\sigma(\bm{r})$. Under the local density approximation, the Lee-Huang-Yang energy density $\varepsilon_{\mathrm{LHY}}[n_\sigma(\bm{r})]$ retains the same functional form as that of Eq.~(\ref{S8}). The total energy correction is then given by
	\begin{align}
		E_\mathrm{LHY} = \int d\bm{r}\, \varepsilon_{\mathrm{LHY}}(\bm{r}).\tag{S9}
	\end{align}

	\vspace{0.5cm}
	\noindent
	
	\section{Analytical Phase Boundaries  in the Homogeneous Case}
	
	\renewcommand{\thefigure}{S1} 
	\begin{figure}[htbp]
		\centering
		\includegraphics[width=0.8\linewidth]{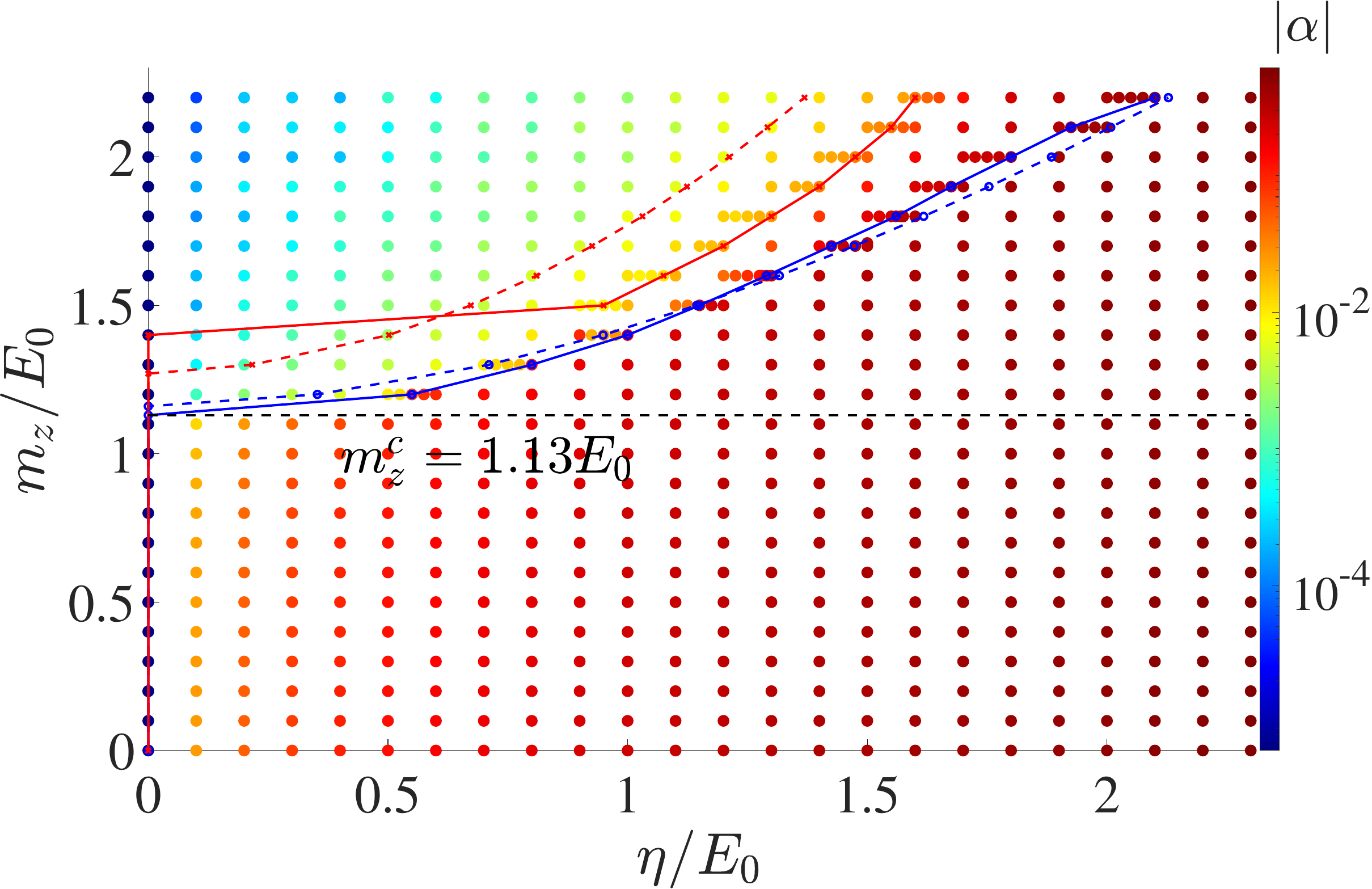}
		\caption{
			Thermal map of the steady-state cavity field $|\alpha|$ as a function of $m_z$ and $\eta$.
			The red solid (dashed) line indicates the numerical (analytical) superradiant phase transition boundary,
			while the blue solid (dashed) line marks the numerical (analytical) gas-liquid transition.
			The intercepts are $m_z^c = 1.13 E_0$ (numerical) and $m_z^c = 1.14 E_0$ (analytical).
			The discrete points in the thermal map correspond to the data points shown in Fig.~\ref{fig:fig3}(a).
		}
		\label{fig:supp}
	\end{figure}
	\renewcommand{\thefigure}{\arabic{figure}} 
	In Fig.~\ref{fig:supp}, the dashed lines show the phase boundaries obtained through analytical calculations based on Eq.~(\ref{Eg}) in the homogeneous case. The calculation parameters for the analytical method are identical to those in Fig.~\ref{fig:fig3}(a), except for the absence of the trapping potential.
	
\end{document}